\documentclass[journal=jpclcd,manuscript=article,layout=twocolumn]{achemso}

\makeatletter
\let\l@addto@macro\relax
\makeatother
\usepackage[fontsize=10pt]{scrextend}
\usepackage{floatflt}

\usepackage{type1cm}
\usepackage{lettrine} 

\usepackage{graphicx}
\usepackage{color}
\usepackage{dcolumn}
\newcolumntype{.}{D{.}{.}{-1}}
\AtBeginDocument{\usepackage{booktabs}}
\usepackage{enumitem,amsmath,amssymb,amsfonts,xfrac}

\usepackage{hyperref}

\newcommand{\GM}{$\overline{\Gamma \text{M}}$}

\newcommand{\GX}{$\overline{\Gamma \text{X}}$}
\newcommand\mc[1]{\multicolumn{1}{c|}{#1}}

\let\oldmaketitle\maketitle
\let\maketitle\relax

\author{G. Benedek}
\affiliation{Donostia International Physics Center (DIPC),~Paseo Manuel de Lardiz{{a}}bal,~4,~20018 Donostia-San Sebastian, Spain}
\affiliation{Dipartimento di Scienza dei Materiali,~Universit{\`a} di Milano-Bicocca,~Via Cozzi 53,~20125 Milano, Italy}
\author{S. Miret-Art{\'e}s}
\email{s.miret@iff.csic.es}
\affiliation{Instituto de F\'isica Fundamental, Consejo Superior de Investigaciones Cient\'ificas, Serrano 123, 28006 Madrid, Spain}
\author{J. R. Manson}
\affiliation{Department of Physics and Astronomy, Clemson University, Clemson, South Carolina 29634, USA}
\author{A. Ruckhofer}
\affiliation{Institute of Experimental Physics, Graz University of Technology, Graz, Austria}
\author{W. E. Ernst}
\affiliation{Institute of Experimental Physics, Graz University of Technology, Graz, Austria}
\author{A. Tamt\"{o}gl}
\email{tamtoegl@gmail.com}
\affiliation{Institute of Experimental Physics, Graz University of Technology, Graz, Austria}

\title[Electron-Phonon Interaction of Topological Semimetal Surfaces]{Origin of the Electron-Phonon Interaction of Topological Semimetal Surfaces Measured with Helium Scattering}

\date{\today}

\begin{document}

\twocolumn[
\begin{@twocolumnfalse}
\oldmaketitle
{\begin{floatingfigure}[r]{0.4\linewidth}
\includegraphics[width=0.37\linewidth]{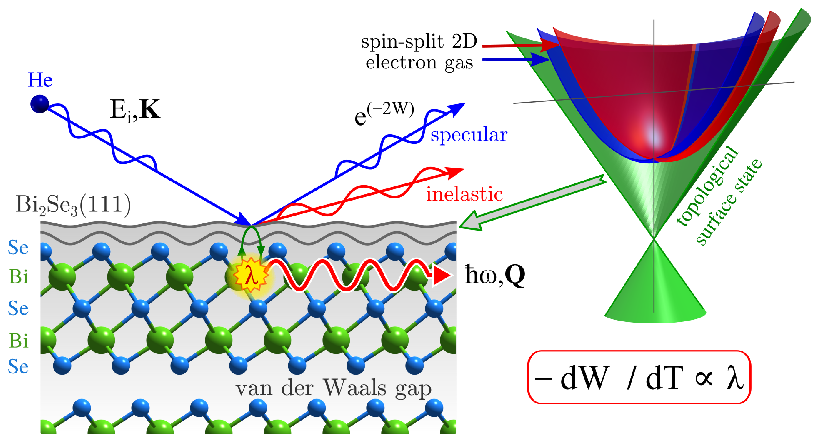}
\end{floatingfigure}
\textbf{ABSTRACT:} He atom scattering has been demonstrated to be a sensitive probe of the electron-phonon interaction parameter $\lambda$ at metal and metal-overlayer surfaces. Here it is shown that the theory linking $\lambda$ to the thermal attenuation of atom scattering spectra (the Debye-Waller factor), can be applied to topological semimetal surfaces, like the quasi-one dimensional charge-density-wave system Bi(114) and the layered pnictogen chalcogenides.\\}
\end{@twocolumnfalse}]

\lettrine[lines=2,nindent=0em]{\textcolor{blue}K}{nowledge} of the electron-phonon (e-ph) interaction at conducting surfaces and the specific role of dimensionality are of great relevance both from a fundamental point of view as well as for various applications, such as two-dimensional (2D)\cite{Saito-17} and quasi-1D superconductivity\cite{WZhang-18} in nanotechnology. Similarly, the e-ph interaction plays a relevant role in other transport properties, e.g., thermoelectricity, in low-dimensional systems like layered Bi and Sb chalcogenides\cite{Liang2016}, and in quasi-crystalline materials which are often viewed as periodic solids in higher dimensions.\cite{Dol09}

In a series of recent experimental and theoretical works, it was shown that the e-ph coupling constants for individual phonons $\lambda_{{\bf Q}, \nu}$, as well as their average $\lambda$ (also known as mass-enhancement parameter or factor),\cite{McMillan-68,Grimvall,Allen} can be measured directly with He-atom scattering (HAS)\cite{Skl,Benedek-14,Manson-JPCL-16,JPCL2,Manson-SurfSciRep}. In particular, the study of multilayer metallic structures\cite{Skl,JPCL2} has shown that HAS can detect subsurface phonons as deep as those that contribute to the e-ph interaction. For example, HAS can detect phonons spanning as many as 10 atomic layers in Pb films\cite{Skl,Benedek-14} (known as the {\em quantum sonar effect}), thus providing the individual $\lambda_{{\bf Q}, \nu}$ values for phonons which provide the dominant contributions to $\lambda$. The values of $\lambda$ are obtained directly from the temperature dependence of the HAS Debye-Waller (DW) exponent, and the interaction range can be assessed from the number of layers, $n_{sat}$, above which the measured $\lambda$ becomes thickness-independent. In that analysis, the conducting surface region of a 3D material could be viewed as a stack of (interacting) 2D electron gases (2DEGs), allowing for the simpler formalism characterizing the 2DEG.\cite{JPCL2} Due to the appreciable depth explored by the e-ph interaction, the values of $\lambda$ obtained from HAS (hereafter called $\lambda_{HAS}$) generally are close to the most reliable values found in the literature,\cite{Manson-JPCL-16,JPCL2} thus allowing one to assess the validity of the new method.

In this work we investigate the specific role of dimensionality in the e-ph mass-enhancement factor $\lambda_{HAS}$ as derived from HAS. The method is shown to be particularly suitable for different classes of conducting 2D materials, such as layered chalcogenides, topological insulators, and systems characterized by a quasi-1D free electron gas, including Bi(114). The present analysis shows that the charge density wave (CDW) transition in Bi(114), recently observed with HAS\cite{Hofmann2019}, is sustained by multi-valley e-ph interaction with a pronounced 1D character. In the case of topological materials, the present analysis of previous HAS data on Bi$_2$Te$_3$(111)\cite{Tam17} and  Bi$_2$Se$_3$(111)\cite{Ruc19} as well as new experimental data on Bi$_2$Te$_2$Se(111) indicates the overwhelming contribution to $\lambda_{HAS}$ from the surface quantum well states as compared to that of the Dirac states.

The DW factor describes the attenuation due to the thermal atomic motion of the elastically scattered intensity $I(T)$ observed at temperature $T$, with respect to the elastic intensity of the corresponding rigid surface $I_0$. It is a multiplicative factor usually written as an exponential function, $\exp\{ -2 W({\bf k}_f, {\bf k}_i ,T)  \}$, of the final (${\bf k}_f$) and incident (${\bf k}_i$) wavevectors of the scattered atom, i.e.,
\begin{equation} \label{Eq1-1}
 I(T) ~=~ I_0 e^{-2W(T)}   
 ~,
\end{equation}
where it is implicit that all quantities in Eq.~(\ref{Eq1-1}) depend on the scattering wavevectors $({\bf k}_f, {\bf k}_i )$. For a two-body collision model, where the incident atom directly interacts with the surface target, the DW exponent is simply expressed by $2 W({\bf k}_f, {\bf k}_i ,T) = \left\langle (\Delta {\bf k}\cdot {\bf u} )^2  \right\rangle_T$, where $\Delta {\bf k} = ( {\bf k}_f - {\bf k}_i )$ is the scattering vector, ${\bf u}$ is the phonon displacement experienced by the projectile atom upon collision, and $ \left\langle \cdot \cdot \cdot \right\rangle_T$  indicates a thermal average. However, atoms incident on a conducting surface with energies generally well below 100 meV are scattered exclusively by the surface free-electron density, a few~\AA~away from the first atomic layer, so that the exchange of energy with the phonon gas only occurs via the phonon-induced modulation of the surface free-electron gas, i.e., via the e-ph interaction. Therefore, it is logical that $2 W({\bf k}_f, {\bf k}_i ,T) $, which originates from the integrated action of all phonons weighted by their respective Bose factors, turns out to be directly proportional, under reasonable approximations, to the mass-enhancement factor $\lambda$.

The expression of $\lambda_{HAS}$ derived in Ref. \cite{Manson-JPCL-16,Manson-JPCL-16-Corr} for a 3D free-electron gas is readily extended to any dimension $d$:
\begin{equation}
\lambda_{HAS}^{(d)} ~=~ - \frac{\phi \gamma_d}{(k_F r_0)^d} ~\frac{k_F^2}{k_{iz}^2} ~
\frac{\partial \ln \{ I\left( T\right)\} }{k_B~ \partial T}
~,
\label{eq:d1}
\end{equation}
where $r_0$ is a lattice distance ($r_0^2\equiv A_c$ for the surface unit cell area in 2D, $r_0^3\equiv V_c$ the unit cell volume in 3D), $\phi$ the workfunction, $k_F$ the Fermi wavevector, $k_{iz}$ is the perpendicular component of the incident wave vector, $k_B$ is the Boltzmann constant, $I(T)$ is the diffraction peak intensity, and $\gamma_d  ~\equiv~ 2^{d-1} ~\pi^{d/2} ~ \Gamma\left(\frac{d}{2}\right)$,\cite{Coxeter} with $\Gamma$ the Gamma-function. As mentioned above and discussed in Ref. \cite{JPCL2} , the 3DEG of a thick slab can be viewed as a pile of $n_{sat}$ 2DEGs, where $n_{sat}$ is the number above the one at which HAS reflectivity becomes independent of thickness. This yields the definition $n_{sat}  ~=~ c^* k_{F\perp}/\pi$, where $k_{F\perp}$ is the Fermi wavevector normal to the surface and $c^*$ is the e-ph interaction range normal to the surface, i.e., the maximum depth beneath the surface from where phonon displacements can modulate the surface charge density. Note that $\pi / k_{F\perp}$ is the wavelength of a Fermi-level charge density $\cos^2{k_{F\perp}z}$, i.e., the nominal thickness of a single 2DEG.

In this way the 2D expression of the e-ph coupling constant for a three-dimensional crystal is obtained\cite{JPCL2}, which is reproduced here for the special case of measurements at the specular condition:
\begin{equation} 
\lambda_{HAS}^{(2D)} ~=~ \frac{\pi}{2 n_{sat}} \alpha,  ~~~~\alpha ~\equiv~ -\frac{\phi}{A_c k^2_{iz}} \frac{\partial \ln\{I(T)\}}{k_B \partial T}
~.
\label{2d3x}
\end{equation}

When applying Eq.~(\ref{2d3x}), it is important to distinguish between metallic surfaces, which present to the He atoms a soft repulsive potential plus a weak long-range attractive well, and layered semimetal surfaces, where the free electron gas is protected by an anion surface layer that results in a hard-wall potential plus a comparatively deep attractive van der Waals potential. In the latter case, $k_{iz}^2$   needs to be corrected due to  the acceleration that the He atom undergoes when entering the attractive well, before being repelled by the hard wall (Beeby correction\cite{Beeby}). This is made with the substitution $k_{iz}^2 \longrightarrow k_{iz}^2 + 2mD/\hbar^2$, where $m$ is the He atom mass and $D$ is the attractive potential depth (generally derived from He-surface bound-state resonances). In many experiments, the incident energy $E_i$  is generally much larger than $D$, so the Beeby correction may be neglected, but not, for example, in $^3$He spin-echo experiments, where $E_i$ is low and comparable to $D$.\cite{Jardine2009}

Low-dimensional free electron gases are often characterized by a CDW instability below a critical temperature $T_c$, generally induced by e-ph interaction via the Fr\"ohlich-Peierls\cite{Peierls,Frolich} or the Kelly-Falicov multivalley mechanisms; \cite{Falicov-1,Falicov-2,Falicov-3} the former typically applying to metal surfaces with a CDW wavevector corresponding to some nesting wavevector at the Fermi contours, the latter more appropriate to semimetal surfaces with pocket states at the Fermi level\cite{Tam19}. The phonon-induced transitions between narrow pockets (nests) realize what is meant as perfect nesting. Since He atoms scattered from a conducting surface probe the surface charge density directly, the occurrence of a CDW below $T_c$ yields additional $T$-dependent diffraction peaks in the elastic scattering angular distribution at parallel wavevector transfers $\Delta K = | \Delta {\bf K} |$ equal or close to the nesting vectors $Q_c$ (i.e., $Q_c = 2k_F$ for the 1D Peierls mechanism). It should be noted that the high sensitivity of HAS permits the detection of weak surface CDWs that are difficult to detect with other methods. An interesting question is whether the temperature dependence of the CDW diffraction peaks carries additional information on the e-ph interaction which sustains the CDW transition.

\begin{figure*}[htb]
\begin{minipage}{\linewidth}
\centering
\includegraphics[width=0.85\textwidth]{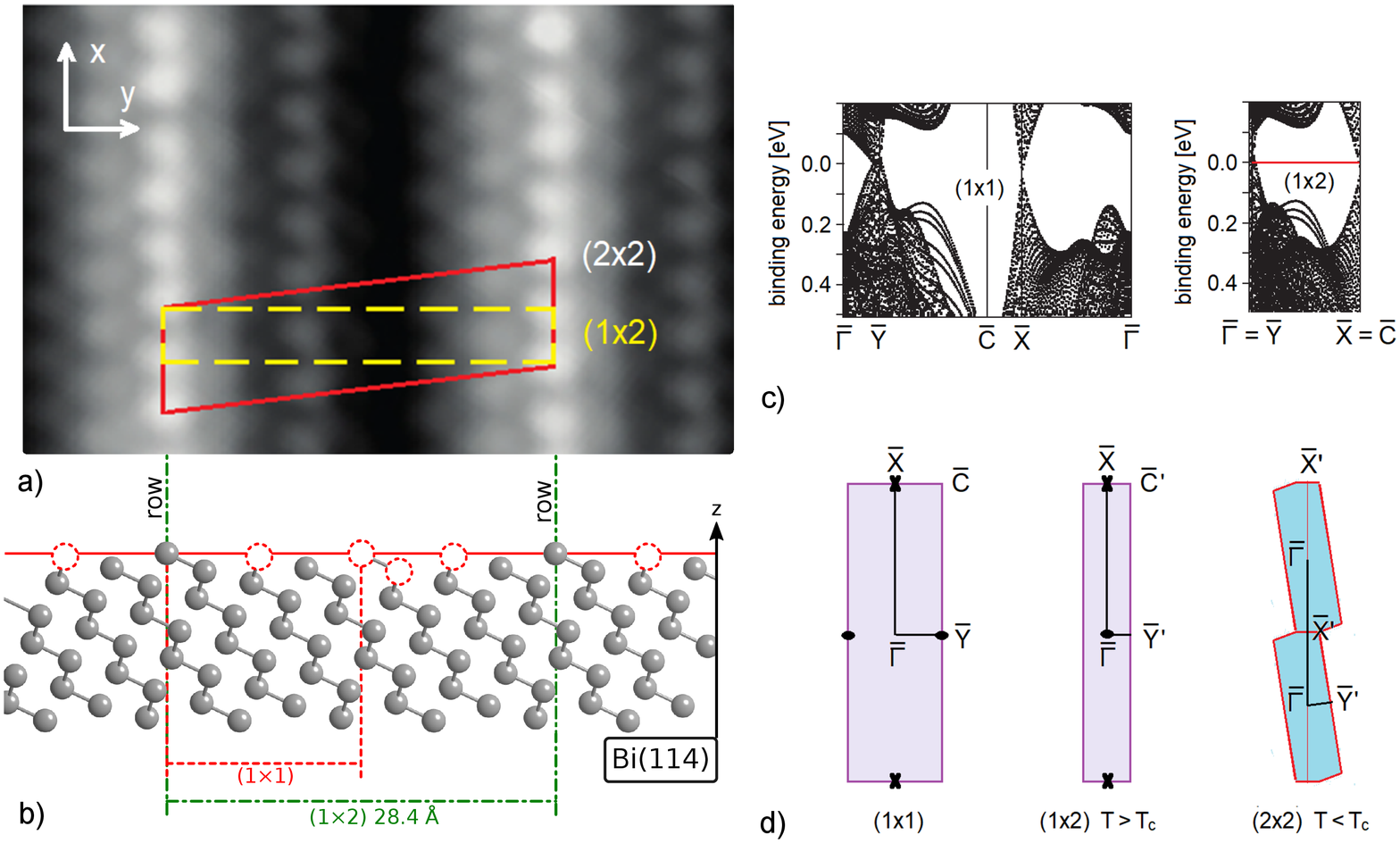}
\caption[]{a) The Bi(114) surface is characterized by parallel atomic rows due to a $(1\times2)$ surface reconstruction as visualised in STM images\cite{Wells}\footnote{\label{note1}\url{https://doi.org/10.1103/PhysRevLett.102.096802}, reproduced with permission, copyright 2009 by the American Physical Society}. b) A side view of the ideal (114) truncation of bismuth (including the red dashed circles) and the actual $(1\times2)$ surface reconstruction (circles removed), giving rise to the parallel atomic rows with an inter-row distance of $28.4$~\AA~and an interatomic distance of $4.54$~\AA. The corresponding (1$\times$2) electronic structure c) is schematically represented as a folding of the calculated electronic structure \cite{Wells}\textsuperscript{\ref{note1}} for the truncated Bi(114) (1$\times$1) surface, with the corresponding surface BZ shown in d). In particular, the cones of electronic states occurring at the Fermi level at the $\overline{\text{X}}$  and $\overline{\text{Y}}$ symmetry points turn out to be aligned in the \GX\  direction after the (1$\times$2) folding [panel d)]. It allows for a multivalley 1D CDW instability along the rows leading to a (2$\times$2) dimerization below $\approx 280$ K and a corresponding CDW observed with HAS.\cite{Hofmann2019} The (2$\times$2) surface portion reproduced in a)\cite{Wells} shows a $\pi$-dephasing of two adjacent rows so as to give a rhombohedric cell, with the corresponding BZ shown in d).}
\label{fig:Bi114}
\end{minipage}
\end{figure*}

When considering the temperature dependence of a diffraction peak intensity for a wavevector transfer $\Delta {\bf K}$ equal to either a ${\bf G}$-vector of the unreconstructed surface lattice $(\Delta {\bf K}  = {\bf G}) $, or to a CDW wavevector $Q_c$, the DW exponent also involves the longitudinal mean-square phonon displacement. For an isotropic mean-square displacement, Eq.~(\ref{eq:d1}) also can be applied to diffraction peaks by replacing $4k_{iz}^2$ with $\Delta k_{z}^2 + \Delta {\bf K}^2$, calculated at the actual scattering geometry at which the diffraction peak is observed. In most HAS experiments the condition $\Delta {\bf K}^2 << \Delta k_{z}^2$ holds, so little difference is expected between the $T$-dependence of the diffraction and specular peaks, provided $\lambda_{HAS}$ is independent, as it should be, of the scattering channel chosen in the experiment.

There is, however, a caveat for the use of a CDW diffraction intensity  $I_{CDW}(T)$. In Eq.~(\ref{Eq1-1}) it has been assumed implicitly that $W(T)$ includes all the temperature dependence of $I(T)$ and that this originates exclusively from thermal vibrations. This is clearly not true for the diffraction from a surface CDW which forms below $T_c$ from a Fermi surface instability and has the temperature-dependent population of electron states near the Fermi level according to Fermi statistics. In this case $I_0$ has an implicit dependence on $T$, which generally is negligible with respect to that of $W(T)$, except near $T_c$: here its square root $\sqrt{I_0}$ works as an order parameter,\cite{order,order-2} and vanishes for increasing $T \to T_c$ as $(1-T/T_c)^\beta$, where $\beta$  is the order-parameter critical exponent (typically $\beta = 1/3$\cite{Liu,Liu-2,Hofmann2019,Old-2H-Ta-paper}, while $T_c \approx 280$ K in the present case\cite{Hofmann2019}). 

\begin{figure*}[htb]
\centering
\includegraphics[width=0.95\textwidth]{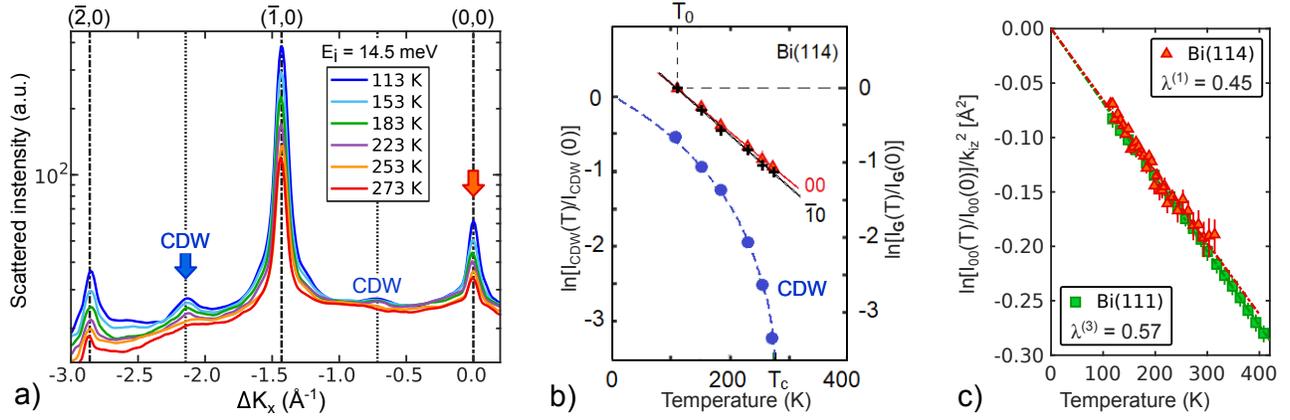}
\caption{Helium atom scattering data from Bi(114): a) HAS angular distributions for several different temperatures, ranging from 113 to 273 K as marked, showing both diffraction peaks of the $(2 \times 1)$ reconstruction and the appearance of the CDW feature according to a $(2 \times 2)$ superstructure. b) The temperature dependence of the $(\sfrac{-3}{2},0)$ CDW peak (left ordinate scale) and of the $(0,0)$ (specular) and $(\overline{1},0)$ (diffraction) peak DW exponents referred to the lowest temperature measured, T = 113 K. c) The DW exponents of the Bi(114) and Bi(111) specular peaks, when divided by the respective squared perpendicular wavevector transfers, show similar slopes but lead to different e-ph coupling strengths.}
\label{fig:Bi114-2}
\end{figure*}

As a good 1DEG example, it is shown that a CDW diffraction peak also may be used to extract $\lambda_{HAS}$ away from the critical region. The ideal (114) truncation of bismuth (\autoref{fig:Bi114}) is characterized by parallel atomic rows along the $x \equiv [1 \overline{1} 0]$  direction, separated by $7.1$~\AA~in the normal direction  $y \equiv [2 2 \overline{1}]$,  with a unit cell including two rows ($b = 14.2$~\AA) and one atom per row (atom spacing along the rows $a = 4.54$~\AA). At room temperature, the Bi(114) surface is reconstructed in a  $(1 \times 2)$ fashion with 3 missing rows out of 4, so as to have one row per unit cell ($b = 28.4$~\AA) and one atom per row  [\autoref{fig:Bi114}a) and~b)]. The electronic structure, calculated by Wells {\em et al.}\cite{Wells} for the $(1 \times 1)$ phase [\autoref{fig:Bi114}c), left], shows cones centered at the $\overline{\text{X}}$ and $\overline{\text{Y}}$ points [\autoref{fig:Bi114}c) and~d)] at the Fermi level. Those at $\overline{\text{Y}}$ are folded into $\overline{\Gamma}$ in the $(1 \times 2)$ reconstructed phase. Both electronic structures allow for a multivalley CDW via e-ph interaction, the former with a 2D character, the latter with a pure 1D character due to the cone alignment along \GX\ with a $G/2$ spacing. HAS angular distributions along \GX\cite{Hofmann2019} [\autoref{fig:Bi114-2}a)] show the growth of additional peaks at $\pm G/2$ and $\pm 3G/2$ below $ T \approx 280$ K,  indicating the formation of a surface commensurate CDW.\cite{Hofmann2019} The associated $(2 \times 2)$ reconstruction consists of a dimerisation along the rows. The portion of the Bi(114)-$(2 \times 2)$ STM image reproduced in \autoref{fig:Bi114}b) from  Hofmann {\em et al.}\cite{Hofmann,Hofmann2019} suggests a phase correlation between rows, giving an oblique ($2 \times 2$) unit cell and a corresponding elongated hexagonal Brillouin zone (BZ) [\autoref{fig:Bi114}d)].

The HAS DW exponents for the specular ${\bf G} = (0,0)$, diffraction $ {\bf G} = (\overline{1},0)$, and CDW $(\sfrac{-3}{2},0)$ peaks measured as a function of temperature below $T_c$ are plotted in \autoref{fig:Bi114-2}b). The specular and diffraction DW exponents have almost the same slopes, the small difference being compensated by the ratio $(\Delta k_{iz}^2)_{(0,0)}/ [\Delta k_z^2 + \Delta {\bf K}^2]_{(\overline{1}, 0)}$, resulting in the same values of $\lambda_{HAS}^{1D}$  within less than 1\%. The input data in Eq.~(\ref{eq:d1}) for $d = 1$ are $\phi = 4.23$ eV\cite{k}, $2k_F = G/2 = 0.7$~\AA$^{-1}$, $r_0(2 \times 2) = 9.08$~\AA, incident energy $E_i = 14.5$ meV, and fixed scattering angle of $91.5^{\circ}$, so $(\Delta k_{iz}^2)_{(0,0)}= 54.3 $~\AA$^{-2}$. The  $(\overline{1},0)$ diffraction occurs at the incident angle of $51.2^{\circ}$, which gives $(\Delta k_{iz}^2)_{(\overline{1},0)}= 53.8$~\AA$^{-2}$, and the resulting e-ph coupling constant is $\lambda_{HAS}^{1D} = 0.45 \pm 0.03$, being the same value for both specular and diffractive channels. 
        
The CDW $(\sfrac{-3}{2},0)$ peak intensity, \autoref{fig:Bi114-2}b), shows the expected critical behavior with $\beta \cong 1/3$, so that a value of $\lambda_{HAS}^{1D}$ can only be estimated from the slope at the lowest temperatures. This is smaller than that for the specular peak by $\sim 5\%$  and is compensated for approximately the same amount by the correcting factor $ (\Delta k_{iz}^2)_{(0,0)}/ [\Delta k_z^2 + \Delta {\bf K}^2]_{({3/2}, 0)} = 1.047$, the incident angle for the CDW peak at $(\sfrac{-3}{2},0)$ being $62.75^\circ$. Thus it is reasonable to conclude that consistent values of $\lambda_{HAS}^{1D}$  can be extracted from the $T$-dependence of the CDW peaks.

It is interesting to compare the value $\lambda_{HAS}^{1D}  = 0.45$ for Bi(114) to that previously derived for Bi(111), either treated as a 3D system where $\lambda_{HAS}^{3D} = 0.57$\cite{Manson-JPCL-16}, in agreement with the value of $\lambda = 0.60$ in Hofmann's review\cite{Hofmann}, or as a 2D system with $n_s = 2$ (a single bilayer) where it is found that $\lambda_{HAS}^{2D}  = 0.40$,\cite{Manson-SurfSciRep} in fair agreement with a recent ab-initio calculation by Ortigoza {\em et al.} for Bi(111)\cite{Ortizoga-14} which yielded $\lambda = 0.45$, just as found here for Bi(114). As seen in \autoref{fig:Bi114-2}c), the DW exponent has about the same slope for Bi(111) and Bi(114), when it is divided by $k_{iz}^2$, in order to account for the different incident energy used in HAS experiments. The fact that $\lambda_{HAS}^{(1D)}$ [Bi(114)] $<$  $\lambda_{HAS}^{(3D)}$[Bi(111)] reflects the dimensionality effect of $\gamma_d$, in the prefactor of Eq.~(\ref{eq:d1}). Incidentally, we note that treating Bi(114) as a 2D system would yield a five times smaller, probably unphysical, value for $\lambda_{HAS}$  due to the large surface unit cell area.

Layered chalcogenides, such as 2D topological materials, with strong intralayer and weak interlayer forces form a wide class of quasi-2D materials with a conducting surface. Some transition-metal dichalcogenides (TMDC) have been investigated with HAS since the late eighties in connection with CDW transitions, related Kohn anomalies in the bulk, and surface phonon dispersion curves.\cite{Ben87,Ben88,Bru89,Bru90,Ben92,Ben94} More recently, HAS studies have been extended to the surface of other TMDCs like 2H-MoS$_2$(0001) \cite{Ane19a} and 1T-PtTe$_2$ \cite{Ane19b}, as well as to pnictogen chalcogenides with surface topological electronic bands at the Fermi level, like Bi$_2$Te$_3$ \cite{How13, Tam17,Tam18}, Bi$_2$Se$_3$ \cite{Zhu11,Ruc19}, etc. The 2D expression for the e-ph coupling constant $\lambda^{(2D)}_{HAS}$ in Eq. (\ref{2d3x}) is the one to be used for these systems. When dealing with the e-ph coupling constant $\lambda_{HAS}$ expressed as an average over the whole phonon spectrum and over all electronic transitions across the Fermi level, natural questions are: a) which phonons contribute most, and b) which electronic states at the Fermi level are more important.

\begin{figure*}[htb]
\begin{minipage}{\linewidth}
\centering
\includegraphics[width=0.55\linewidth]{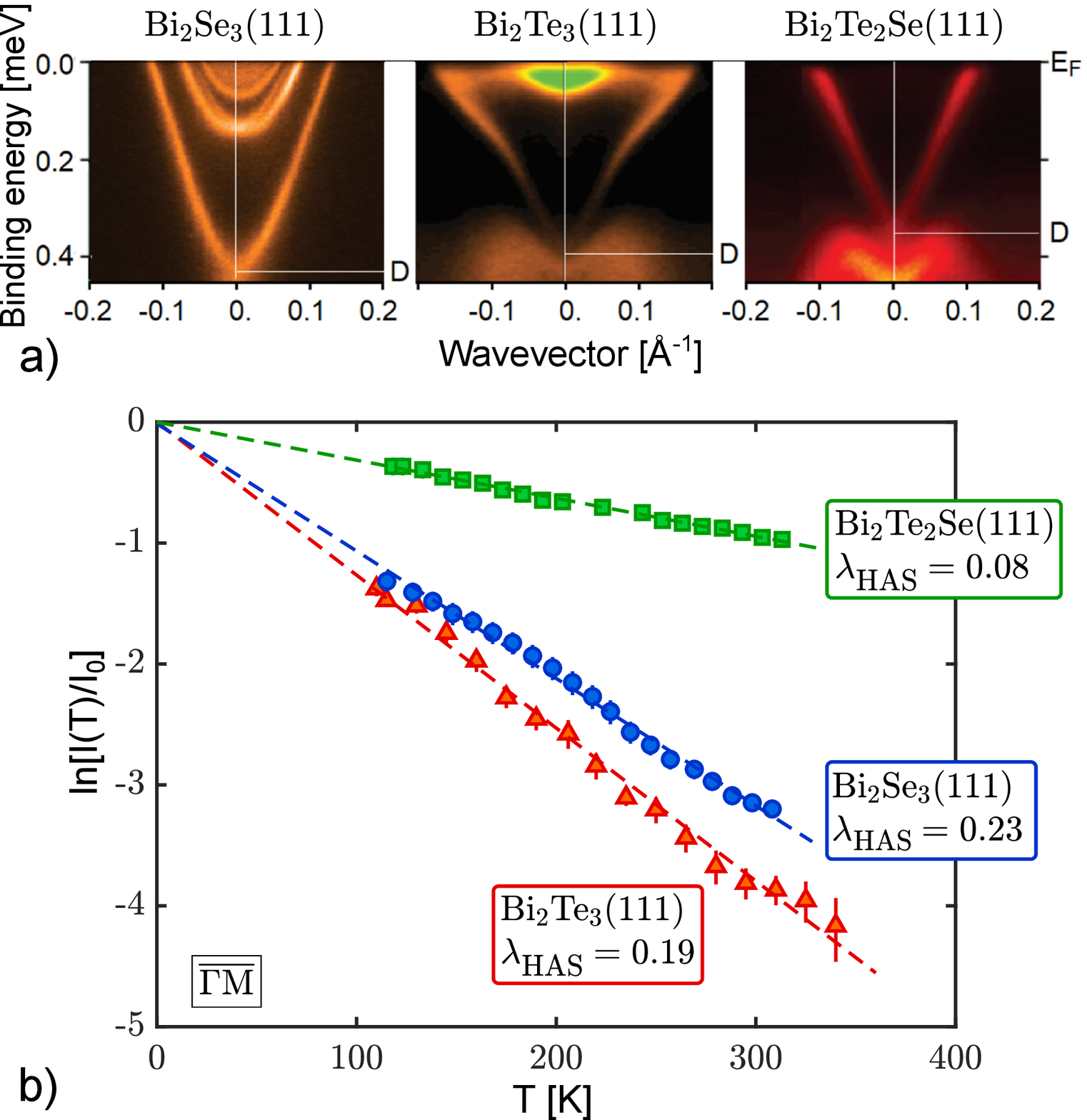}
\caption[]{Three different Bi chalcogenides Bi$_2$Se$_3$, Bi$_2$Te$_3$ and Bi$_2$Te$_2$Se. Top panels: ARPES data for the (111) surface of three Bi chalcogenides with decreasing binding energy of the Dirac point (D) and of the surface conduction-band minimum (from $0.15$ eV in Bi$_2$Se$_3$ \cite{Bia12}\footnote{\url{https://arxiv.org/abs/1206.1563}} to $0.08$ eV in Bi$_2$Te$_3$ \cite{Mic14}\footnote{\url{https://arxiv.org/abs/1403.3050}}, and $\approx 0$ in Bi$_2$Te$_2$Se \cite{Bar14}\footnote{\url{https://doi.org/10.1021/nl501489m}, reproduced with permission, copyright 2014 by the American Chemical Society.}). Bottom panel, the DW exponent slopes from HAS specular intensity measured as functions of temperature with the scattering plane in the \GM\ direction for the same samples. The corresponding e-ph coupling constants $\lambda_{HAS}$ decrease from Bi$_2$Se$_3$(111) to Bi$_2$Te$_2$Se(111), suggesting a dominant role in the e-ph interaction of the conduction band quantum-well electronic states over the Dirac electrons.}
\label{fig:Bi2chalc}
\end{minipage}
\end{figure*}

The theoretical analysis by Heid {\em et al.} \cite{Hei17} of the mode-selected e-ph coupling constants $\lambda_{{\bf Q}\nu}$ shows that in pnictogen chalcogenides, optical phonons give the major contribution to e-ph interaction, and therefore to the DW exponent. Both Bi$_2$Se$_3$(111)\cite{Ruc19} and Bi$_2$Te$_3$(111)\cite{Tam18} exhibit two highly dispersed optical branches with deep minima at $\overline{\Gamma}$ for 3\textsuperscript{rd}-layer longitudinal polarization and at $\approx$\GM$/2$ for (mostly) 3\textsuperscript{rd}-layer shear-vertical (SV3) polarisation. Their optical character and largest amplitude at the central chalcogen layer of the quintuple layer endow these modes with a dipolar character and therefore a large e-ph interaction, consistent with the Heid {\em et al.} theoretical analysis \cite{Hei17}. Spin-echo $^3$He scattering data from Bi$_2$Te$_3$(111) \cite{Tam18} suggest a Kohn anomaly also in the longitudinal acoustic branch corresponding to a nesting across the Dirac cone above the surface conduction-band minimum. As discussed in \cite{Hei17}, the interband e-ph coupling occurring when the Fermi level is above the surface conduction-band minimum is enhanced largely by the involvement of surface quantum-well states. This conclusion is confirmed by the following analysis of  $\lambda^{(2D)}_{HAS}$ in Bi chalcogenides as a function of the Fermi level position.

\begin{table*}[htb]
\centering
\caption{Input data for the calculation from the HAS DW exponent, and results for the e-ph coupling constant $\lambda^{(2D)}_{HAS}$ (with a relative uncertainty of about 10\%) and comparison with values from other sources.\\}
\label{tab:EPhTab}
\resizebox{\textwidth}{!}{
\begin{tabular}{ c | . | . | c | . | . | . | . | . }
\toprule    
Surface  & \mc{$ k^2_{iz}$}  & \mc{ $ \phi $ } & \mc{$\lambda_{TF}$} & \mc{ $A_{c}$ } & \mc{$c_0$} &\mc{$D$} &  \mc{$\lambda^{(2D)}_{HAS}$} & \multicolumn{1}{c}{$\lambda$} \\
& \mc{ $[\mbox{\AA}^{-2}]$ }  & \mc{ [eV] } & \mc{[$\mbox{\AA}$]}  & \mc{[$\mbox{\AA}^2$]} & \mc{ [$\mbox{\AA}$] }& \mc{ [meV] } & \mc{ } & \multicolumn{1}{c}{ (other Refs)}  \\[1mm]
       \midrule
       \rule{0mm}{5mm}
        & & & & & & & & 0.17 \cite{Che13} \\ 
Bi$_2$Se$_3$(111) & 10.1 \cite{Ruc19} & 4.9 \cite{Suh14} & $\approx60$ \cite{Bia10}& 14.92 & 9.60 & 6.54 \cite{Ruc19a}
& 0.23 & 0.25 \cite{Hat11} \\
& & & & & & & & 0.26 \cite{Zel15} \\[1mm]
\hline
\rule{0mm}{5mm}
Bi$_2$Te$_3$(111)  & 9.9 \cite{Tam17} & 4.9  \cite{Suh14}&  $\approx100$ \cite{Pet13} &16.46  & 10.16 & 6.22\cite{Tam18a}& 0.19  & 0.19\cite{Che13} \\[1mm]
\hline
\rule{0mm}{5mm}
Bi$_2$Te$_{2}$Se(111) & 10.71\textsuperscript{\ref{note2}} & 4.9\cite{Suh14} & $\approx100$ \cite{Geh12}  & 16.09 & 10.0 
& 6.4\textsuperscript{\ref{note3}} & 0.08 & 0.12 \cite{Che13} \\[1mm] 
\bottomrule
\end{tabular}}
\begin{enumerate}[label=(\alph*)]
\setlength\itemsep{0em}
\item This work\label{note2}
\item Average over Bi$_2$Se$_3$(111) and Bi$_2$Te$_3$(111)\label{note3}
\end{enumerate}
\end{table*}

The temperature dependence of HAS reflectivity from the three Bi chalcogenide surfaces  Bi$_2$Se$_3$(111), Bi$_2$Te$_3$(111), and Bi$_2$Te$_{3-x}$Se$_x$(111) (phase II with $x \approx  1$ \cite{Mi13}$^{,}$\footnote{According to the surface lattice constant $a = 4.31~\mbox{\AA}$ as determined by HAS and Fig. 1(b) of \cite{Mi13}, $x \approx  1$ for the present sample.}, hereafter approximated by Bi$_2$Te$_2$Se(111)) has been measured for three samples whose surface electronic states near the Fermi level, in particular the topological Dirac states and the quantum-well states above the surface conduction-band minimum, are known from angle-resolved photoemission spectroscopy (ARPES) data \cite{Bia10,Bia12,Mic14,Bar14}. As seen in \autoref{fig:Bi2chalc} (top panel), the binding energy of the Dirac point (D) with respect to the Fermi energy $E_F$ decreases in the sequence Bi$_2$Se$_3$(111), Bi$_2$Te$_3$(111), Bi$_2$Te$_2$Se(111), as does the surface conduction-band minimum (from $0.15$ eV, to $0.08$ eV and $\approx 0$, respectively). Correspondingly, the DW exponent slope derived from the HAS specular intensity as a function of temperature also decreases. Similar behavior is expected for the e-ph coupling constant $\lambda^{(2D)}_{HAS}$, which is shown in \autoref{tab:EPhTab}. The latter is derived from Eq. (\ref{2d3x}) by setting  $n_{sat} = 2 \lambda_{TF} / c_0$, where $c_0$ is the quintuple layer (QL) thickness and $\lambda_{TF}$ is the Thomas-Fermi screening length, accounting for the surface band-bending extension in degenerate semiconductors and semimetals \cite{Shk84}. The factor 2 in the above expression of $n_{sat}$  accounts for the fact that each QL contains two metal (Bi) layers. Note that the Fermi-level density of states includes the factor 2 for spin multiplicity, and this is appropriate for the quantum-well states above the surface conduction-band minimum that mostly contribute to $\lambda^{(2D)}_{HAS}$. On the other hand, no factor of 2 in $n_{sat}$ is necessary when only the Dirac states are involved, due to their multiplicity of 1. The uncertainties ($\pm$) given for $\lambda_{HAS}$ in the following are based on the confidence bounds of the DW slope. Other sources in terms of the uncertainty are $A_c$, $\phi$ and $\lambda_{TF}$ with the largest contribution likely to be due to $\lambda_{TF}$. Taken together it is safe to assume a relative uncertainty of about 10\% for $\lambda_{HAS}$ as shown in \autoref{tab:EPhTab}.

With the input data collected in \autoref{tab:EPhTab} and the DW exponent slopes displayed in \autoref{fig:Bi2chalc} (bottom panel), it is found  that $\lambda^{(2D)}_{HAS}= 0.23\pm 0.01$ for Bi$_2$Se$_3$(111), $0.19\pm 0.01$ for Bi$_2$Te$_3$(111), and $0.080\pm0.004$ for Bi$_2$Te$_2$Se(111), in good agreement with selected results from other sources (\autoref{tab:EPhTab}, last column). This clearly indicates the dominant role of surface quantum well (QW) states over the modest contribution of Dirac electrons. The enhancement effect of QW states and related interband transitions has been investigated thoroughly by Chen {\em et al.} \cite{Che13} with high resolution ARPES for the family Bi$_2$Te$_{3-x}$Se$_x$(111) $(0 \leq  x \leq 3)$, including n-type  Bi$_2$Te$_3$(111), and theoretically for  Bi$_2$Se$_3$(111) and Bi$_2$Te$_3$(111) by Heid {\em et al.} \cite{Hei17}. As shown by Pan {\em et al.} \cite{Pan12}, in accurate ARPES studies on bulk Bi$_2$Se$_3$(111) samples, where only Dirac topological states are involved because the Fermi level is less than 0.3 eV above the Dirac point, the e-ph coupling constant turns out to be quite small, ranging from $0.076 \pm 0.007$ to $0.088 \pm 0.009$, similar to that found with HAS in Bi$_2$Te$_2$Se(111) under similar conditions.

Since the surface QW states extend into the bulk on the order of  $\lambda_{TF}$, i.e., much longer than the penetration of surface Dirac states, it is interesting to compare the above results for  $\lambda^{(2D)}_{HAS}$ with the corresponding values of $\lambda^{(3D)}_{HAS}$ when these materials are treated as 3D materials. The ratio $\lambda^{(3D)}_{HAS} / \lambda^{(2D)}_{HAS} = \pi / (k_F \lambda_{TF})$, with $k_F \simeq 0.1$ $\mbox{\AA}^{-1}$ [\autoref{fig:Bi2chalc} (top)] and $\lambda_{TF}$ representing the 3DEG thickness, turns out to be $\simeq 1$. 

Unlike Bi(114), where the quasi-1D character of the electron gas is quite evident, in layered pnictogen chalcogenides the considerable penetration of the QW states gives  $\lambda^{(2D)}_{HAS} \simeq \lambda^{(3D)}_{HAS}$. This is consistent with the fact that the QW states are the surface states which provide the major contribution to the e-ph interaction. Information about which phonons contribute most to $\lambda$ can also be obtained from inelastic HAS intensities, as explained in the introduction. The Kohn anomalies reported in the lower part of the phonon spectrum  \cite{How13,Tam18,Zhu11} are indicative of a strong e-ph coupling for specific wavevectors and frequencies, though it has been predicted that the major contribution in these materials comes from polar optical modes\cite{Hei17}. Indeed this is in agreement with recent HAS measurements of the phonon dispersion curves in Bi$_2$Se$_3$(111)\cite{Ruc19}, which indicate the longitudinal optical branch L3 (with the largest displacement on the 3\textsuperscript{rd} (Se) atomic plane) as the one having the largest mode-selective e-ph coupling.\\[0.5ex]

In conclusion, it has been shown that the temperature dependence of HAS reflectivity allows for the determination of the electron-phonon coupling constant of topological semimetal surfaces. In the case of the quasi-1D Bi(114) surface, the DW factor from the CDW diffraction peak yields an e-ph coupling constant $\lambda^{(1D)}_{HAS}$ consistent with that derived from the reflectivity. Therefore the e-ph interaction acts as the driving mechanism for the observed multi-valley CDW transition. In the absence of spin-orbit coupling, the phonon angular momentum cannot convert into an electron spin-flip, so no good nesting would be allowed across the Dirac cone, and only the strong spin-orbit coupling occurring in topological materials allows for a comparatively weak e-ph {\it intra}-cone interaction. The multi-valley mechanism at the zone boundary overcomes the nesting problem, because with more Dirac cones separated by less than a {\bf G} vector, there is always a good {\it inter}-cone (i.e., multi-valley) nesting, even for opposite chiralities. Such a favorable circumstance, allowing for a substantial $\lambda^{(1D)}_{HAS}$ in Bi(114) and a CDW transition, does not occur in pnictogen chalcogenides, due to the single Dirac cone location at the center of the BZ. Most of their appreciable e-ph interaction is provided by the QW states, as long as they are located at the Fermi level. The present extension of HAS $\lambda$-analysis from metal surfaces \cite{Benedek-14,Manson-JPCL-16} and thin metal films \cite{JPCL2} to topological semimetal surfaces qualifies He atom scattering as a universal tool for the measurement of electron-phonon coupling in conducting low-dimensional systems.

\section{Experimental Methods}
The experimental data of this work was obtained at the HAS apparatus in Graz\cite{Tam2010} and the $^3$He spin-echo scattering apparatus in Cambridge\cite{Jardine2009}. In both cases the scattered intensity of a nearly monochromatic He beam in the range of $8-15$ meV is monitored as a function of incident angle $\vartheta_i$ and at various surface temperatures. The DW measurement of Bi$_2$Te$_3$ can be found in \cite{Tam17} while the DW data of Bi$_2$Se$_3$ is reported in Ref. \cite{Ruc19}. Most of the Bi(114) data has been published in the work of Hofmann {\em et al.}\cite{Hofmann2019}, whereas the Bi$_2$Te$_2$Se experimental data is presented here for the first time.

\section{Acknowledgements} 
One of us (G.B.) would  like to thank Profs.~M. Bernasconi (Milano-Bicocca) and E. V. Chulkov and P. M. Echenique (DIPC, San Sebastian) for many helpful discussions. This work is partially supported by a grant with Ref. FIS2017-83473-C2-1-P from the Ministerio de Ciencia Universidades e Innovaci\'on (Spain). A.R., W.E.E. and A.T. acknowledge financial support provided by the FWF (Austrian Science Fund) within the projects J3479-N20 and P29641-N36. G. B. gratefully acknowledges the Italian Ministry of University and Research (MIUR) for financial support through grant “Dipartimenti di Eccellenza - 2017 Materials for Energy". We would like to thank Martin Bremholm, Ellen M. J. Hedegaard, and Bo B. Iversen for the synthesis of the samples, Marco Bianchi for his advice and help in terms of the sample preparation, and Philip Hofmann as well as the aforementioned people for many helpful discussions and additional characterizations of the samples.  


\begin{mcitethebibliography}{63}
\providecommand*\natexlab[1]{#1}
\providecommand*\mciteSetBstSublistMode[1]{}
\providecommand*\mciteSetBstMaxWidthForm[2]{}
\providecommand*\mciteBstWouldAddEndPuncttrue
  {\def\EndOfBibitem{\unskip.}}
\providecommand*\mciteBstWouldAddEndPunctfalse
  {\let\EndOfBibitem\relax}
\providecommand*\mciteSetBstMidEndSepPunct[3]{}
\providecommand*\mciteSetBstSublistLabelBeginEnd[3]{}
\providecommand*\EndOfBibitem{}
\mciteSetBstSublistMode{f}
\mciteSetBstMaxWidthForm{subitem}{(\alph{mcitesubitemcount})}
\mciteSetBstSublistLabelBeginEnd
  {\mcitemaxwidthsubitemform\space}
  {\relax}
  {\relax}

\bibitem[Saito \latin{et~al.}(2017)Saito, Nojima, and Iwasa]{Saito-17}
Saito,~Y.; Nojima,~T.; Iwasa,~Y. {Highly crystalline 2D superconductors}.
  \emph{Nat. Rev. Mater} \textbf{2017}, \emph{2}, 16094\relax
\mciteBstWouldAddEndPuncttrue
\mciteSetBstMidEndSepPunct{\mcitedefaultmidpunct}
{\mcitedefaultendpunct}{\mcitedefaultseppunct}\relax
\EndOfBibitem
\bibitem[Zhang and Melo(2018)Zhang, and Melo]{WZhang-18}
Zhang,~W.; Melo,~C. A. R. S.~D. In \emph{{Quasi-One-Dimensional Organic
  Superconductors}}; Zhang,~W., Ed.; Advanced Physics Series; World Scientific,
  2018; Vol.~5\relax
\mciteBstWouldAddEndPuncttrue
\mciteSetBstMidEndSepPunct{\mcitedefaultmidpunct}
{\mcitedefaultendpunct}{\mcitedefaultseppunct}\relax
\EndOfBibitem
\bibitem[Liang \latin{et~al.}(2016)Liang, Cheng, Zhang, Liu, and
  Zhang]{Liang2016}
Liang,~J.; Cheng,~L.; Zhang,~J.; Liu,~H.; Zhang,~Z. {Maximizing the
  thermoelectric performance of topological insulator Bi$_2$Te$_3$ films in the
  few-quintuple layer regime}. \emph{Nanoscale} \textbf{2016}, \emph{8},
  8855--8862\relax
\mciteBstWouldAddEndPuncttrue
\mciteSetBstMidEndSepPunct{\mcitedefaultmidpunct}
{\mcitedefaultendpunct}{\mcitedefaultseppunct}\relax
\EndOfBibitem
\bibitem[Dolin\v{s}ek \latin{et~al.}(2009)Dolin\v{s}ek, Smontara,
  Bari\v{s}i\'{c}, and Gille]{Dol09}
Dolin\v{s}ek,~J.; Smontara,~A.; Bari\v{s}i\'{c},~O.~S.; Gille,~P.
  {Phonon-enhanced thermoelectric power of Y-Al-Ni-Co decagonal approximant}.
  \emph{Z. Kristallogr.} \textbf{2009}, \emph{224}, 64\relax
\mciteBstWouldAddEndPuncttrue
\mciteSetBstMidEndSepPunct{\mcitedefaultmidpunct}
{\mcitedefaultendpunct}{\mcitedefaultseppunct}\relax
\EndOfBibitem
\bibitem[McMillan(1968)]{McMillan-68}
McMillan,~W.~L. {Transition Temperature of Strong-Coupled Superconductors}.
  \emph{Phys. Rev} \textbf{1968}, \emph{167}, 331\relax
\mciteBstWouldAddEndPuncttrue
\mciteSetBstMidEndSepPunct{\mcitedefaultmidpunct}
{\mcitedefaultendpunct}{\mcitedefaultseppunct}\relax
\EndOfBibitem
\bibitem[Grimvall(1981)]{Grimvall}
Grimvall,~G. \emph{{The Electron-Phonon Interaction in Metals}}; North-Holland:
  New York, 1981\relax
\mciteBstWouldAddEndPuncttrue
\mciteSetBstMidEndSepPunct{\mcitedefaultmidpunct}
{\mcitedefaultendpunct}{\mcitedefaultseppunct}\relax
\EndOfBibitem
\bibitem[Allen(1972)]{Allen}
Allen,~P.~B. {Neutron Spectroscopy of Superconductors}. \emph{Phys. Rev. B}
  \textbf{1972}, \emph{6}, 2577\relax
\mciteBstWouldAddEndPuncttrue
\mciteSetBstMidEndSepPunct{\mcitedefaultmidpunct}
{\mcitedefaultendpunct}{\mcitedefaultseppunct}\relax
\EndOfBibitem
\bibitem[Sklyadneva \latin{et~al.}(2011)Sklyadneva, Benedek, Chulkov,
  Echenique, Heid, Bohnen, and Toennies]{Skl}
Sklyadneva,~I.~Y.; Benedek,~G.; Chulkov,~E.~V.; Echenique,~P.~M.; Heid,~R.;
  Bohnen,~K.-P.; Toennies,~J.~P. {Mode-Selected Electron-Phonon Coupling in
  Superconducting Pb Nanofilms Determined from He Atom Scattering}. \emph{Phys.
  Rev. Lett.} \textbf{2011}, \emph{107}, 095502\relax
\mciteBstWouldAddEndPuncttrue
\mciteSetBstMidEndSepPunct{\mcitedefaultmidpunct}
{\mcitedefaultendpunct}{\mcitedefaultseppunct}\relax
\EndOfBibitem
\bibitem[Benedek \latin{et~al.}(2014)Benedek, Bernasconi, Bohnen, Campi,
  Chulkov, Echenique, Heid, Sklyadneva, and Toennies]{Benedek-14}
Benedek,~G.; Bernasconi,~M.; Bohnen,~K.-P.; Campi,~D.; Chulkov,~E.~V.;
  Echenique,~P.~M.; Heid,~R.; Sklyadneva,~I.~Y.; Toennies,~J.~P. {Unveiling
  mode-selected electron-phonon interactions in metal films by helium atom
  scattering}. \emph{Phys. Chem. Chem. Phys} \textbf{2014}, \emph{16},
  7159\relax
\mciteBstWouldAddEndPuncttrue
\mciteSetBstMidEndSepPunct{\mcitedefaultmidpunct}
{\mcitedefaultendpunct}{\mcitedefaultseppunct}\relax
\EndOfBibitem
\bibitem[Manson \latin{et~al.}(2016)Manson, Benedek, and
  Miret-Art\'es]{Manson-JPCL-16}
Manson,~J.~R.; Benedek,~G.; Miret-Art\'es,~S. {Electron-phonon coupling
  strength at metal surfaces directly determined from the helium atom
  scattering Debye-Waller factor}. \emph{J. Phys. Chem. Lett.} \textbf{2016},
  \emph{7}, 1016\relax
\mciteBstWouldAddEndPuncttrue
\mciteSetBstMidEndSepPunct{\mcitedefaultmidpunct}
{\mcitedefaultendpunct}{\mcitedefaultseppunct}\relax
\EndOfBibitem
\bibitem[Benedek \latin{et~al.}(2018)Benedek, Miret-Art\'es, Toennies, and
  Manson]{JPCL2}
Benedek,~G.; Miret-Art\'es,~S.,~S.; Toennies,~J.~P.; Manson,~J.~R. {The
  Electron–Phonon Coupling Constant of Metallic Overlayers from Specular He
  Atom Scattering}. \emph{J. Phys. Chem. Lett.} \textbf{2018}, \emph{9},
  76--83\relax
\mciteBstWouldAddEndPuncttrue
\mciteSetBstMidEndSepPunct{\mcitedefaultmidpunct}
{\mcitedefaultendpunct}{\mcitedefaultseppunct}\relax
\EndOfBibitem
\bibitem[Manson \latin{et~al.}(unpublished)Manson, Benedek, and
  Miret-Art\'es]{Manson-SurfSciRep}
Manson,~J.~R.; Benedek,~G.; Miret-Art\'es,~S. {Atom scattering as a probe of
  the surface electron-phonon interaction}. \emph{Surf. Sci. Rep.}
  \textbf{unpublished}, \relax
\mciteBstWouldAddEndPunctfalse
\mciteSetBstMidEndSepPunct{\mcitedefaultmidpunct}
{}{\mcitedefaultseppunct}\relax
\EndOfBibitem
\bibitem[Hofmann \latin{et~al.}(2019)Hofmann, Ugeda, Tamt\"ogl, Ruckhofer,
  Ernst, Benedek, Mart\'{\i}nez-Galera, Str\'o\ifmmode~\dot{z}\else
  \.{z}\fi{}ecka, G\'omez-Rodr\'{\i}guez, Rienks, Jensen, Pascual, and
  Wells]{Hofmann2019}
Hofmann,~P.; Ugeda,~M.~M.; Tamt\"ogl,~A.; Ruckhofer,~A.; Ernst,~W.~E.;
  Benedek,~G.; Mart\'{\i}nez-Galera,~A.~J.; Str\'o\ifmmode~\dot{z}\else
  \.{z}\fi{}ecka,~A.; G\'omez-Rodr\'{\i}guez,~J.~M.; Rienks,~E. \latin{et~al.}
  {Strong-coupling charge density wave in a one-dimensional topological metal}.
  \emph{Phys. Rev. B} \textbf{2019}, \emph{99}, 035438\relax
\mciteBstWouldAddEndPuncttrue
\mciteSetBstMidEndSepPunct{\mcitedefaultmidpunct}
{\mcitedefaultendpunct}{\mcitedefaultseppunct}\relax
\EndOfBibitem
\bibitem[Tamt\"ogl \latin{et~al.}(2017)Tamt\"ogl, Kraus, Avidor, Bremholm,
  Hedegaard, Iversen, Bianchi, Hofmann, Ellis, Allison, Benedek, and
  Ernst]{Tam17}
Tamt\"ogl,~A.; Kraus,~P.; Avidor,~N.; Bremholm,~M.; Hedegaard,~E. M.~J.;
  Iversen,~B.~B.; Bianchi,~M.; Hofmann,~P.; Ellis,~J.; Allison,~W.
  \latin{et~al.}  {Electron-Phonon Coupling and Surface Debye Temperature of
  Bi$_2$Te$_3$(111) from Helium Atom Scattering}. \emph{Phys. Rev. B}
  \textbf{2017}, \emph{95}, 195401\relax
\mciteBstWouldAddEndPuncttrue
\mciteSetBstMidEndSepPunct{\mcitedefaultmidpunct}
{\mcitedefaultendpunct}{\mcitedefaultseppunct}\relax
\EndOfBibitem
\bibitem[Ruckhofer \latin{et~al.}(2020)Ruckhofer, Campi, Bremholm, Hofmann,
  Benedek, Bernasconi, Ernst, and Tamt\"ogl]{Ruc19}
Ruckhofer,~A.; Campi,~D.; Bremholm,~M.; Hofmann,~P.; Benedek,~G.;
  Bernasconi,~M.; Ernst,~W.~E.; Tamt\"ogl,~A. {THz Surface Modes and
  Electron-Phonon Coupling in Bi$_2$Se$_3$(111)}. \emph{Phys. Rev. Res.}
  \textbf{2020}, \emph{in review}, preprint,
  \url{https://arxiv.org/abs/1907.01864}\relax
\mciteBstWouldAddEndPuncttrue
\mciteSetBstMidEndSepPunct{\mcitedefaultmidpunct}
{\mcitedefaultendpunct}{\mcitedefaultseppunct}\relax
\EndOfBibitem
\bibitem[Manson \latin{et~al.}(2016)Manson, Benedek, and
  Miret-Artés]{Manson-JPCL-16-Corr}
Manson,~J.~R.; Benedek,~G.; Miret-Artés,~S. {Correction to “Electron-Phonon
  Coupling Strength at Metal Surfaces Directly Determined from the Helium Atom
  Scattering Debye-Waller Factor”}. \emph{J. Phys. Chem. Lett.}
  \textbf{2016}, \emph{7}, 1691--1691\relax
\mciteBstWouldAddEndPuncttrue
\mciteSetBstMidEndSepPunct{\mcitedefaultmidpunct}
{\mcitedefaultendpunct}{\mcitedefaultseppunct}\relax
\EndOfBibitem
\bibitem[Coxeter(1969)]{Coxeter}
Coxeter,~H. S.~M. \emph{{Introduction to Geometry}}, second edition ed.; John
  Wiley and Sons: New York, 1969\relax
\mciteBstWouldAddEndPuncttrue
\mciteSetBstMidEndSepPunct{\mcitedefaultmidpunct}
{\mcitedefaultendpunct}{\mcitedefaultseppunct}\relax
\EndOfBibitem
\bibitem[Beeby(1971)]{Beeby}
Beeby,~J.~L. {Scattering of Helium Atoms from Surfaces}. \emph{J. Physics C}
  \textbf{1971}, \emph{4}, L359\relax
\mciteBstWouldAddEndPuncttrue
\mciteSetBstMidEndSepPunct{\mcitedefaultmidpunct}
{\mcitedefaultendpunct}{\mcitedefaultseppunct}\relax
\EndOfBibitem
\bibitem[Jardine \latin{et~al.}(2009)Jardine, Hedgeland, Alexandrowicz,
  Allison, and Ellis]{Jardine2009}
Jardine,~A.; Hedgeland,~H.; Alexandrowicz,~G.; Allison,~W.; Ellis,~J. {Helium-3
  spin-echo: principles and application to dynamics at surfaces}. \emph{Prog.
  Surf. Sci.} \textbf{2009}, \emph{84}, 323--379\relax
\mciteBstWouldAddEndPuncttrue
\mciteSetBstMidEndSepPunct{\mcitedefaultmidpunct}
{\mcitedefaultendpunct}{\mcitedefaultseppunct}\relax
\EndOfBibitem
\bibitem[Peierls(1955)]{Peierls}
Peierls,~R.~E. \emph{{Quantum theory of solids}}; Oxford Univ. Press: Oxford,
  1955\relax
\mciteBstWouldAddEndPuncttrue
\mciteSetBstMidEndSepPunct{\mcitedefaultmidpunct}
{\mcitedefaultendpunct}{\mcitedefaultseppunct}\relax
\EndOfBibitem
\bibitem[Fr\"ohlich(1954)]{Frolich}
Fr\"ohlich,~H. {On the Theory of Superconductivity: The One-Dimensional Case}.
  \emph{Proc. R. Soc. Lond} \textbf{1954}, \emph{223}, 296\relax
\mciteBstWouldAddEndPuncttrue
\mciteSetBstMidEndSepPunct{\mcitedefaultmidpunct}
{\mcitedefaultendpunct}{\mcitedefaultseppunct}\relax
\EndOfBibitem
\bibitem[Kelly and Falicov(1977)Kelly, and Falicov]{Falicov-1}
Kelly,~M.~J.; Falicov,~L.~M. {Electronic Ground-State of Inversion Layers in
  Many-Valley Semiconductors}. \emph{Phys. Rev. B} \textbf{1977}, \emph{15},
  1974\relax
\mciteBstWouldAddEndPuncttrue
\mciteSetBstMidEndSepPunct{\mcitedefaultmidpunct}
{\mcitedefaultendpunct}{\mcitedefaultseppunct}\relax
\EndOfBibitem
\bibitem[Kelly and Falicov(1977)Kelly, and Falicov]{Falicov-2}
Kelly,~M.~J.; Falicov,~L.~M. {Optical-Properties of Charge-Density-Wave
  ground-States for Inversion Layers in Many-Valley Semiconductors}.
  \emph{Phys. Rev. B} \textbf{1977}, \emph{15}, 1983\relax
\mciteBstWouldAddEndPuncttrue
\mciteSetBstMidEndSepPunct{\mcitedefaultmidpunct}
{\mcitedefaultendpunct}{\mcitedefaultseppunct}\relax
\EndOfBibitem
\bibitem[Kelly and Falicov(1976)Kelly, and Falicov]{Falicov-3}
Kelly,~M.~J.; Falicov,~L.~M. {Electronic Structure of Inversion Layers in
  Many-Valley Semiconductors}. \emph{Phys. Rev. Lett.} \textbf{1976},
  \emph{37}, 1021\relax
\mciteBstWouldAddEndPuncttrue
\mciteSetBstMidEndSepPunct{\mcitedefaultmidpunct}
{\mcitedefaultendpunct}{\mcitedefaultseppunct}\relax
\EndOfBibitem
\bibitem[Tamt\"ogl \latin{et~al.}(2019)Tamt\"ogl, Kraus,
  Mayrhofer-Reinhartshuber, Benedek, Bernasconi, Dragoni, Campi, and
  Ernst]{Tam19}
Tamt\"ogl,~A.; Kraus,~P.; Mayrhofer-Reinhartshuber,~M.; Benedek,~G.;
  Bernasconi,~M.; Dragoni,~D.; Campi,~D.; Ernst,~W.~E. {Statics and Dynamics of
  Multivalley Charge Density Waves in Sb(111)}. \emph{npj Quantum Mater.}
  \textbf{2019}, \emph{4}, 18\relax
\mciteBstWouldAddEndPuncttrue
\mciteSetBstMidEndSepPunct{\mcitedefaultmidpunct}
{\mcitedefaultendpunct}{\mcitedefaultseppunct}\relax
\EndOfBibitem
\bibitem[Wells \latin{et~al.}(2009)Wells, Dil, Meier, Lobo-Checa, Petrov,
  Osterwalder, Ugeda, Fernandez-Torrente, Pascual, Rienks, Jensen, and
  Hofmann]{Wells}
Wells,~J.~W.; Dil,~J.~H.; Meier,~F.; Lobo-Checa,~J.; Petrov,~V.~N.;
  Osterwalder,~J.; Ugeda,~M.~M.; Fernandez-Torrente,~I.; Pascual,~J.~I.;
  Rienks,~E. D.~L. \latin{et~al.}  {Nondegenerate Metallic States on Bi(114): A
  One-Dimensional Topological Metal}. \emph{Phys. Rev. Lett.} \textbf{2009},
  \emph{102}, 096802\relax
\mciteBstWouldAddEndPuncttrue
\mciteSetBstMidEndSepPunct{\mcitedefaultmidpunct}
{\mcitedefaultendpunct}{\mcitedefaultseppunct}\relax
\EndOfBibitem
\bibitem[Liu \latin{et~al.}(2019)Liu, Ma, Wang, and Yang]{order}
Liu,~R.; Ma,~T.; Wang,~S.; Yang,~J. {Thermodynamical Potentials of Classical
  and Quantum Systems}. \emph{Discrete and Continuous Dynamical Systems-Series
  B} \textbf{2019}, \emph{24}, 1411\relax
\mciteBstWouldAddEndPuncttrue
\mciteSetBstMidEndSepPunct{\mcitedefaultmidpunct}
{\mcitedefaultendpunct}{\mcitedefaultseppunct}\relax
\EndOfBibitem
\bibitem[Ma and Wang(2009)Ma, and Wang]{order-2}
Ma,~T.; Wang,~S. {Cahn-Hilliard Equations and Phase Transition Dynamics for
  Binary Systems}. \emph{Discrete and Continuous Dynamical Systems B}
  \textbf{2009}, \emph{11}, 741\relax
\mciteBstWouldAddEndPuncttrue
\mciteSetBstMidEndSepPunct{\mcitedefaultmidpunct}
{\mcitedefaultendpunct}{\mcitedefaultseppunct}\relax
\EndOfBibitem
\bibitem[Ge and Liu(2013)Ge, and Liu]{Liu}
Ge,~Y.; Liu,~A.~Y. {Phonon-mediated superconductivity in electron-doped
  single-layer MoS$_2$: A first-principles prediction}. \emph{Phys. Rev. B}
  \textbf{2013}, \emph{87}, 241408\relax
\mciteBstWouldAddEndPuncttrue
\mciteSetBstMidEndSepPunct{\mcitedefaultmidpunct}
{\mcitedefaultendpunct}{\mcitedefaultseppunct}\relax
\EndOfBibitem
\bibitem[Liu(2009)]{Liu-2}
Liu,~A.~Y. {Electron-phonon coupling in compressed 1T-TaS$_2$: Stability and
  superconductivity from first principles}. \emph{Phys. Rev. B} \textbf{2009},
  \emph{79}, 220515\relax
\mciteBstWouldAddEndPuncttrue
\mciteSetBstMidEndSepPunct{\mcitedefaultmidpunct}
{\mcitedefaultendpunct}{\mcitedefaultseppunct}\relax
\EndOfBibitem
\bibitem[Benedek \latin{et~al.}(1994)Benedek, Hofmann, Ruggerone, Onida, and
  Miglio]{Old-2H-Ta-paper}
Benedek,~G.; Hofmann,~F.; Ruggerone,~P.; Onida,~G.; Miglio,~L. {Surface Phonons
  in Layered Crystals - Theoretical Aspects}. \emph{Surf. Sci. Rep.}
  \textbf{1994}, \emph{20}, 3\relax
\mciteBstWouldAddEndPuncttrue
\mciteSetBstMidEndSepPunct{\mcitedefaultmidpunct}
{\mcitedefaultendpunct}{\mcitedefaultseppunct}\relax
\EndOfBibitem
\bibitem[Hofmann(2006)]{Hofmann}
Hofmann,~P. {The surfaces of bismuth: Structural and electronic properties}.
  \emph{Prog. Surf. Sci.} \textbf{2006}, \emph{81}, 191\relax
\mciteBstWouldAddEndPuncttrue
\mciteSetBstMidEndSepPunct{\mcitedefaultmidpunct}
{\mcitedefaultendpunct}{\mcitedefaultseppunct}\relax
\EndOfBibitem
\bibitem[Michaelson(1977)]{k}
Michaelson,~H.~B. {The work function of the elements and its periodicity}.
  \emph{J. Appl. Phys.} \textbf{1977}, \emph{48}, 4729\relax
\mciteBstWouldAddEndPuncttrue
\mciteSetBstMidEndSepPunct{\mcitedefaultmidpunct}
{\mcitedefaultendpunct}{\mcitedefaultseppunct}\relax
\EndOfBibitem
\bibitem[Ortigoza \latin{et~al.}(2014)Ortigoza, Sklyadneva, Heid, Chulkov,
  Rahman, Bohnen, and Echenique]{Ortizoga-14}
Ortigoza,~M.~A.; Sklyadneva,~I.~Y.; Heid,~R.; Chulkov,~E.~V.; Rahman,~T.~S.;
  Bohnen,~K.-P.; Echenique,~P.~M. {Ab initio lattice dynamics and
  electron-phonon coupling of Bi(111)}. \emph{Phys. Rev. B} \textbf{2014},
  \emph{90}, 195438\relax
\mciteBstWouldAddEndPuncttrue
\mciteSetBstMidEndSepPunct{\mcitedefaultmidpunct}
{\mcitedefaultendpunct}{\mcitedefaultseppunct}\relax
\EndOfBibitem
\bibitem[Benedek \latin{et~al.}(1987)Benedek, Miglio, Skofronick, Brusdeylins,
  Heimlich, and Toennies]{Ben87}
Benedek,~G.; Miglio,~L.; Skofronick,~J.~G.; Brusdeylins,~G.; Heimlich,~C.;
  Toennies,~J.~P. {Surface Phonon Dynamics in 2H-TaSe$_2$(001)}. \emph{J. Vac.
  Sci Technol. A} \textbf{1987}, \emph{5}, 1093\relax
\mciteBstWouldAddEndPuncttrue
\mciteSetBstMidEndSepPunct{\mcitedefaultmidpunct}
{\mcitedefaultendpunct}{\mcitedefaultseppunct}\relax
\EndOfBibitem
\bibitem[Benedek \latin{et~al.}(1988)Benedek, Brusdeylins, Heimlich, Miglio,
  Skofronick, and Toennies]{Ben88}
Benedek,~G.; Brusdeylins,~G.; Heimlich,~C.; Miglio,~L.; Skofronick,~J.;
  Toennies,~J.~P. {Shifted Surface Phonon Anomaly in 2H-TaSe$_2$(001)}.
  \emph{Phys. Rev. Lett.} \textbf{1988}, \emph{60}, 1037\relax
\mciteBstWouldAddEndPuncttrue
\mciteSetBstMidEndSepPunct{\mcitedefaultmidpunct}
{\mcitedefaultendpunct}{\mcitedefaultseppunct}\relax
\EndOfBibitem
\bibitem[Brusdeylins \latin{et~al.}(1989)Brusdeylins, Heimlich, Skofronick,
  Toennies, Vollmer, and Benedek]{Bru89}
Brusdeylins,~G.; Heimlich,~C.; Skofronick,~J.~G.; Toennies,~J.~P.; Vollmer,~R.;
  Benedek,~G. {Determination of the Critical Exponents for a Charge Density
  Wave Transition in 2H-TaSe$_2$ by Helium Atom Scattering}. \emph{Euro Phys
  Lett} \textbf{1989}, \emph{9}, 563\relax
\mciteBstWouldAddEndPuncttrue
\mciteSetBstMidEndSepPunct{\mcitedefaultmidpunct}
{\mcitedefaultendpunct}{\mcitedefaultseppunct}\relax
\EndOfBibitem
\bibitem[Brusdeylins \latin{et~al.}(1990)Brusdeylins, Hofmann, Toennies,
  Vollmer, Benedek, Ruggerone, and Skofronick]{Bru90}
Brusdeylins,~G.; Hofmann,~F.; Toennies,~J.~P.; Vollmer,~R.; Benedek,~G.;
  Ruggerone,~P.; Skofronick,~J. In \emph{Phonons}; S.~Hunklinger,~W.~L.,
  Weiss,~G., Eds.; World Sci.: Singapore, 1990; p 892\relax
\mciteBstWouldAddEndPuncttrue
\mciteSetBstMidEndSepPunct{\mcitedefaultmidpunct}
{\mcitedefaultendpunct}{\mcitedefaultseppunct}\relax
\EndOfBibitem
\bibitem[Benedek \latin{et~al.}(1992)Benedek, Miglio, and Seriani]{Ben92}
Benedek,~G.; Miglio,~L.; Seriani,~G. In \emph{Helium Atom Scattering from
  Surfaces}; Hulpke,~E., Ed.; Springer: Heidelberg, 1992; pp 207--242\relax
\mciteBstWouldAddEndPuncttrue
\mciteSetBstMidEndSepPunct{\mcitedefaultmidpunct}
{\mcitedefaultendpunct}{\mcitedefaultseppunct}\relax
\EndOfBibitem
\bibitem[Benedek \latin{et~al.}(1994)Benedek, Brusdeylins, Hofmann, Ruggerone,
  Toennies, Vollmer, and Skofronick]{Ben94}
Benedek,~G.; Brusdeylins,~G.; Hofmann,~F.; Ruggerone,~P.; Toennies,~J.;
  Vollmer,~R.; Skofronick,~J. {Strong coupling of Rayleigh phonons to charge
  density waves in 1T-TaS$_2$}. \emph{Surf. Sci.} \textbf{1994}, \emph{304},
  185 -- 190\relax
\mciteBstWouldAddEndPuncttrue
\mciteSetBstMidEndSepPunct{\mcitedefaultmidpunct}
{\mcitedefaultendpunct}{\mcitedefaultseppunct}\relax
\EndOfBibitem
\bibitem[Anemone \latin{et~al.}(2019)Anemone, Taleb, Benedek,
  Castellanos-Gomez, and Far{\'\i}as]{Ane19a}
Anemone,~G.; Taleb,~A.~A.; Benedek,~G.; Castellanos-Gomez,~A.; Far{\'\i}as,~D.
  {Electron-Phonon Coupling Constant of 2H-MoS$_2$(0001) from Helium-Atom
  Scattering}. \emph{J. Phys. Chem. C} \textbf{2019}, \emph{123},
  3682--3686\relax
\mciteBstWouldAddEndPuncttrue
\mciteSetBstMidEndSepPunct{\mcitedefaultmidpunct}
{\mcitedefaultendpunct}{\mcitedefaultseppunct}\relax
\EndOfBibitem
\bibitem[Anemone \latin{et~al.}(2020)Anemone, Garnica, Zappia, Aguilar, Taleb,
  Kuo, Lue, Politano, Benedek, de~Parga, Miranda, and Far{\'{\i}}as]{Ane19b}
Anemone,~G.; Garnica,~M.; Zappia,~M.; Aguilar,~P.~C.; Taleb,~A.~A.; Kuo,~C.-N.;
  Lue,~C.~S.; Politano,~A.; Benedek,~G.; de~Parga,~A. L.~V. \latin{et~al.}
  {Experimental determination of surface thermal expansion and electron-phonon
  coupling constant of 1T-PtTe$_2$}. \emph{2D Mater.} \textbf{2020}, \emph{7},
  025007\relax
\mciteBstWouldAddEndPuncttrue
\mciteSetBstMidEndSepPunct{\mcitedefaultmidpunct}
{\mcitedefaultendpunct}{\mcitedefaultseppunct}\relax
\EndOfBibitem
\bibitem[Howard \latin{et~al.}(2013)Howard, El-Batanouny, Sankar, and
  Chou]{How13}
Howard,~C.; El-Batanouny,~M.; Sankar,~R.; Chou,~F.~C. {Anomalous behavior in
  the phonon dispersion of the (001) surface of Bi$_2$Te$_3$ determined from
  helium atom-surface scattering measurements}. \emph{Phys. Rev. B}
  \textbf{2013}, \emph{88}, 035402\relax
\mciteBstWouldAddEndPuncttrue
\mciteSetBstMidEndSepPunct{\mcitedefaultmidpunct}
{\mcitedefaultendpunct}{\mcitedefaultseppunct}\relax
\EndOfBibitem
\bibitem[Tamt\"ogl \latin{et~al.}(2018)Tamt\"ogl, Campi, Bremholm, Hedegaard,
  Iversen, Bianchi, Hofmann, Marzari, Benedek, Ellis, and Allison]{Tam18}
Tamt\"ogl,~A.; Campi,~D.; Bremholm,~M.; Hedegaard,~E. M.~J.; Iversen,~B.~B.;
  Bianchi,~M.; Hofmann,~P.; Marzari,~N.; Benedek,~G.; Ellis,~J. \latin{et~al.}
  {Nanoscale surface dynamics of Bi$_2$Te$_3$(111): observation of a prominent
  surface acoustic wave and the role of van der Waals interactions}.
  \emph{Nanoscale} \textbf{2018}, \emph{10}, 14627--14636\relax
\mciteBstWouldAddEndPuncttrue
\mciteSetBstMidEndSepPunct{\mcitedefaultmidpunct}
{\mcitedefaultendpunct}{\mcitedefaultseppunct}\relax
\EndOfBibitem
\bibitem[Zhu \latin{et~al.}(2011)Zhu, Santos, Sankar, Chikara, Howard, Chou,
  Chamon, and El-Batanouny]{Zhu11}
Zhu,~X.; Santos,~L.; Sankar,~R.; Chikara,~S.; Howard,~C.~.; Chou,~F.~C.;
  Chamon,~C.; El-Batanouny,~M. {Interaction of Phonons and Dirac Fermions on
  the Surface of ${\mathrm{Bi}}_{2}{\mathrm{Se}}_{3}$: A Strong Kohn Anomaly}.
  \emph{Phys. Rev. Lett.} \textbf{2011}, \emph{107}, 186102\relax
\mciteBstWouldAddEndPuncttrue
\mciteSetBstMidEndSepPunct{\mcitedefaultmidpunct}
{\mcitedefaultendpunct}{\mcitedefaultseppunct}\relax
\EndOfBibitem
\bibitem[Bianchi \latin{et~al.}(2012)Bianchi, Hatch, Guan, Planke, Mi, Iversen,
  and Hofmann]{Bia12}
Bianchi,~M.; Hatch,~R.~C.; Guan,~D.; Planke,~T.; Mi,~J.; Iversen,~B.~B.;
  Hofmann,~P. {The electronic structure of clean and adsorbate-covered
  Bi$_2$Se$_3$: an angle-resolved photoemission study}. \emph{Semicond. Sci.
  Technol.} \textbf{2012}, \emph{27}, 124001\relax
\mciteBstWouldAddEndPuncttrue
\mciteSetBstMidEndSepPunct{\mcitedefaultmidpunct}
{\mcitedefaultendpunct}{\mcitedefaultseppunct}\relax
\EndOfBibitem
\bibitem[Michiardi \latin{et~al.}(2014)Michiardi, Aguilera, Bianchi,
  de~Carvalho, Ladeira, Teixeira, Soares, Friedrich, Bl\"ugel, and
  Hofmann]{Mic14}
Michiardi,~M.; Aguilera,~I.; Bianchi,~M.; de~Carvalho,~V.~E.; Ladeira,~L.~O.;
  Teixeira,~N.~G.; Soares,~E.~A.; Friedrich,~C.; Bl\"ugel,~S.; Hofmann,~P.
  {Bulk band structure of ${\mathrm{Bi}}_{2}{\mathrm{Te}}_{3}$}. \emph{Phys.
  Rev. B} \textbf{2014}, \emph{90}, 075105\relax
\mciteBstWouldAddEndPuncttrue
\mciteSetBstMidEndSepPunct{\mcitedefaultmidpunct}
{\mcitedefaultendpunct}{\mcitedefaultseppunct}\relax
\EndOfBibitem
\bibitem[Barreto \latin{et~al.}(2014)Barreto, K{\"u}hnemund, Edler, Tegenkamp,
  Mi, Bremholm, Iversen, Frydendahl, Bianchi, and Hofmann]{Bar14}
Barreto,~L.; K{\"u}hnemund,~L.; Edler,~F.; Tegenkamp,~C.; Mi,~J.; Bremholm,~M.;
  Iversen,~B.~B.; Frydendahl,~C.; Bianchi,~M.; Hofmann,~P. {Surface-Dominated
  Transport on a Bulk Topological Insulator}. \emph{Nano Lett.} \textbf{2014},
  \emph{14}, 3755--3760\relax
\mciteBstWouldAddEndPuncttrue
\mciteSetBstMidEndSepPunct{\mcitedefaultmidpunct}
{\mcitedefaultendpunct}{\mcitedefaultseppunct}\relax
\EndOfBibitem
\bibitem[Heid \latin{et~al.}(2017)Heid, Sklyadneva, and Chulkov]{Hei17}
Heid,~R.; Sklyadneva,~I.~Y.; Chulkov,~E.~V. {Electron-phonon coupling in
  topological surface states: The role of polar optical modes}. \emph{Sci.
  Rep.} \textbf{2017}, \emph{7}, 1095\relax
\mciteBstWouldAddEndPuncttrue
\mciteSetBstMidEndSepPunct{\mcitedefaultmidpunct}
{\mcitedefaultendpunct}{\mcitedefaultseppunct}\relax
\EndOfBibitem
\bibitem[Chen \latin{et~al.}(2013)Chen, Xie, Feng, Yi, Liang, He, Mou, He,
  Peng, Liu, Liu, Zhao, Liu, Dong, Zhang, Yu, Wang, Peng, Wang, Zhang, Yang,
  Chen, Xu, and Zhou]{Che13}
Chen,~C.; Xie,~Z.; Feng,~Y.; Yi,~H.; Liang,~A.; He,~S.; Mou,~D.; He,~J.;
  Peng,~Y.; Liu,~X. \latin{et~al.}  {Tunable Dirac fermion dynamics in
  topological insulators}. \emph{Sci. Rep.} \textbf{2013}, \emph{3}, 2411\relax
\mciteBstWouldAddEndPuncttrue
\mciteSetBstMidEndSepPunct{\mcitedefaultmidpunct}
{\mcitedefaultendpunct}{\mcitedefaultseppunct}\relax
\EndOfBibitem
\bibitem[Suh \latin{et~al.}(2014)Suh, Fu, Liu, Furdyna, Yu, Walukiewicz, and
  Wu]{Suh14}
Suh,~J.; Fu,~D.; Liu,~X.; Furdyna,~J.~K.; Yu,~K.~M.; Walukiewicz,~W.; Wu,~J.
  {Fermi-level stabilization in the topological insulators Bi$_2$Se$_3$ and
  Bi$_2$Te$_3$: Origin of the surface electron gas}. \emph{Phys. Rev. B}
  \textbf{2014}, \emph{89}, 115307\relax
\mciteBstWouldAddEndPuncttrue
\mciteSetBstMidEndSepPunct{\mcitedefaultmidpunct}
{\mcitedefaultendpunct}{\mcitedefaultseppunct}\relax
\EndOfBibitem
\bibitem[Bianchi \latin{et~al.}(2010)Bianchi, Guan, Bao, Mi, Iversen, King, and
  Hofmann]{Bia10}
Bianchi,~M.; Guan,~D.; Bao,~S.; Mi,~J.; Iversen,~B.~B.; King,~P.~D.;
  Hofmann,~P. {Coexistence of the topological state and a two-dimensional
  electron gas on the surface of Bi$_2$Se$_3$}. \emph{Nat. Commun.}
  \textbf{2010}, \emph{1}, 128\relax
\mciteBstWouldAddEndPuncttrue
\mciteSetBstMidEndSepPunct{\mcitedefaultmidpunct}
{\mcitedefaultendpunct}{\mcitedefaultseppunct}\relax
\EndOfBibitem
\bibitem[Ruckhofer \latin{et~al.}(2019)Ruckhofer, Tamt\"ogl, Pusterhofer,
  Bremholm, and Ernst]{Ruc19a}
Ruckhofer,~A.; Tamt\"ogl,~A.; Pusterhofer,~M.; Bremholm,~M.; Ernst,~W.~E.
  {Helium-Surface Interaction and Electronic Corrugation of Bi$_2$Se$_3$(111)}.
  \emph{J. Phys. Chem. C} \textbf{2019}, \emph{123}, 17829--17841\relax
\mciteBstWouldAddEndPuncttrue
\mciteSetBstMidEndSepPunct{\mcitedefaultmidpunct}
{\mcitedefaultendpunct}{\mcitedefaultseppunct}\relax
\EndOfBibitem
\bibitem[Hatch \latin{et~al.}(2011)Hatch, Bianchi, Guan, Bao, Mi, Iversen,
  Nilsson, Hornek\ae{}r, and Hofmann]{Hat11}
Hatch,~R.~C.; Bianchi,~M.; Guan,~D.; Bao,~S.; Mi,~J.; Iversen,~B.~B.;
  Nilsson,~L.; Hornek\ae{}r,~L.; Hofmann,~P. {Stability of the
  ${\mathbf{Bi}}_{2}{\mathbf{Se}}_{3}$(111) topological state: Electron-phonon
  and electron-defect scattering}. \emph{Phys. Rev. B} \textbf{2011},
  \emph{83}, 241303\relax
\mciteBstWouldAddEndPuncttrue
\mciteSetBstMidEndSepPunct{\mcitedefaultmidpunct}
{\mcitedefaultendpunct}{\mcitedefaultseppunct}\relax
\EndOfBibitem
\bibitem[Zeljkovic \latin{et~al.}(2015)Zeljkovic, Scipioni, Walkup, Okada,
  Zhou, Sankar, Chang, Wang, Lin, Bansil, Chou, Wang, and Madhavan]{Zel15}
Zeljkovic,~I.; Scipioni,~K.~L.; Walkup,~D.; Okada,~Y.; Zhou,~W.; Sankar,~R.;
  Chang,~G.; Wang,~Y.~J.; Lin,~H.; Bansil,~A. \latin{et~al.}  {Nanoscale
  determination of the mass enhancement factor in the lightly doped bulk
  insulator lead selenide}. \emph{Nat. Commun.} \textbf{2015}, \emph{6},
  6559--\relax
\mciteBstWouldAddEndPuncttrue
\mciteSetBstMidEndSepPunct{\mcitedefaultmidpunct}
{\mcitedefaultendpunct}{\mcitedefaultseppunct}\relax
\EndOfBibitem
\bibitem[Pettes \latin{et~al.}(2013)Pettes, Maassen, Jo, Lundstrom, and
  Shi]{Pet13}
Pettes,~M.~T.; Maassen,~J.; Jo,~I.; Lundstrom,~M.~S.; Shi,~L. {Effects of
  Surface Band Bending and Scattering on Thermoelectric Transport in Suspended
  Bismuth Telluride Nanoplates}. \emph{Nano Lett.} \textbf{2013}, \emph{13},
  5316\relax
\mciteBstWouldAddEndPuncttrue
\mciteSetBstMidEndSepPunct{\mcitedefaultmidpunct}
{\mcitedefaultendpunct}{\mcitedefaultseppunct}\relax
\EndOfBibitem
\bibitem[Tamt\"ogl \latin{et~al.}(2018)Tamt\"ogl, Pusterhofer, Bremholm,
  Hedegaard, Iversen, Hofmann, Ellis, Allison, Miret-Art\'es, and
  Ernst]{Tam18a}
Tamt\"ogl,~A.; Pusterhofer,~M.; Bremholm,~M.; Hedegaard,~E.~M.; Iversen,~B.~B.;
  Hofmann,~P.; Ellis,~J.; Allison,~W.; Miret-Art\'es,~S.; Ernst,~W.~E. {A
  Helium-Surface Interaction Potential of Bi$_2$Te$_3$(111) from
  Ultrahigh-Resolution Spin-Echo Measurements}. \emph{Surf. Sci.}
  \textbf{2018}, \emph{678}, 25--31\relax
\mciteBstWouldAddEndPuncttrue
\mciteSetBstMidEndSepPunct{\mcitedefaultmidpunct}
{\mcitedefaultendpunct}{\mcitedefaultseppunct}\relax
\EndOfBibitem
\bibitem[Gehring \latin{et~al.}(2012)Gehring, Gao, Burghard, and Kern]{Geh12}
Gehring,~P.; Gao,~B.~F.; Burghard,~M.; Kern,~K. {Growth of High-Mobility
  Bi$_2$Te$_2$Se Nanoplatelets on hBN Sheets by van der Waals Epitaxy}.
  \emph{Nano Lett.} \textbf{2012}, \emph{12}, 5137\relax
\mciteBstWouldAddEndPuncttrue
\mciteSetBstMidEndSepPunct{\mcitedefaultmidpunct}
{\mcitedefaultendpunct}{\mcitedefaultseppunct}\relax
\EndOfBibitem
\bibitem[Mi \latin{et~al.}(2013)Mi, Bremholm, Bianchi, Borup, Johnsen,
  S{\o}ndergaard, Guan, Hatch, Hofmann, and Iversen]{Mi13}
Mi,~J.-L.; Bremholm,~M.; Bianchi,~M.; Borup,~K.; Johnsen,~S.;
  S{\o}ndergaard,~M.; Guan,~D.; Hatch,~R.~C.; Hofmann,~P.; Iversen,~B.~B.
  {Phase Separation and Bulk p-n Transition in Single Crystals of
  Bi$_2$Te$_2$Se Topological Insulator}. \emph{Adv. Mater} \textbf{2013},
  \emph{25}, 889--893\relax
\mciteBstWouldAddEndPuncttrue
\mciteSetBstMidEndSepPunct{\mcitedefaultmidpunct}
{\mcitedefaultendpunct}{\mcitedefaultseppunct}\relax
\EndOfBibitem
\bibitem[Shklovskii and Efros(1984)Shklovskii, and Efros]{Shk84}
Shklovskii,~B.~I.; Efros,~A.~E. \emph{{Electronic Properties of Doped
  Semiconductors}}; Springer, 1984\relax
\mciteBstWouldAddEndPuncttrue
\mciteSetBstMidEndSepPunct{\mcitedefaultmidpunct}
{\mcitedefaultendpunct}{\mcitedefaultseppunct}\relax
\EndOfBibitem
\bibitem[Pan \latin{et~al.}(2012)Pan, Fedorov, Gardner, Lee, Chu, and
  Valla]{Pan12}
Pan,~Z.-H.; Fedorov,~A.~V.; Gardner,~D.; Lee,~Y.~S.; Chu,~S.; Valla,~T.
  {Measurement of an Exceptionally Weak Electron-Phonon Coupling on the Surface
  of the Topological Insulator ${\mathrm{Bi}}_{2}{\mathrm{Se}}_{3}$ Using
  Angle-Resolved Photoemission Spectroscopy}. \emph{Phys. Rev. Lett.}
  \textbf{2012}, \emph{108}, 187001\relax
\mciteBstWouldAddEndPuncttrue
\mciteSetBstMidEndSepPunct{\mcitedefaultmidpunct}
{\mcitedefaultendpunct}{\mcitedefaultseppunct}\relax
\EndOfBibitem
\bibitem[Tamt\"ogl \latin{et~al.}(2010)Tamt\"ogl, Mayrhofer-Reinhartshuber,
  Balak, Ernst, and Rieder]{Tam2010}
Tamt\"ogl,~A.; Mayrhofer-Reinhartshuber,~M.; Balak,~N.; Ernst,~W.~E.;
  Rieder,~K.~H. Elastic and inelastic scattering of He atoms from Bi(111).
  \emph{J. Phys. Condens. Matter} \textbf{2010}, \emph{22}, 304019\relax
\mciteBstWouldAddEndPuncttrue
\mciteSetBstMidEndSepPunct{\mcitedefaultmidpunct}
{\mcitedefaultendpunct}{\mcitedefaultseppunct}\relax
\EndOfBibitem
\end{mcitethebibliography}
\providecommand{\latin}[1]{#1}
\providecommand*\mcitethebibliography{\thebibliography}
\csname @ifundefined\endcsname{endmcitethebibliography}
  {\let\endmcitethebibliography\endthebibliography}{}

\end{document}